\documentclass{article}

    \usepackage[preprint]{neurips_2025}

\usepackage[utf8]{inputenc} 
\usepackage[T1]{fontenc}    
\usepackage{hyperref}       
\usepackage{url}            
\usepackage{booktabs}       
\usepackage{amsfonts}       
\usepackage{nicefrac}       
\usepackage{microtype}      
\usepackage{xcolor}         
\usepackage[normalem]{ulem} 

\usepackage{xcolor}         
\usepackage{graphicx}       
\usepackage{tikz}           
\usepackage{epsfig}         
\usepackage{subcaption}     
\usepackage{caption}        
\usepackage{wrapfig}        
\usepackage{color}          
\usepackage{colortbl}       
\usepackage{makecell}
\usepackage{amsmath}
\usepackage{multirow}

\title{PCMamba: Physics-Informed Cross-Modal State Space Model for Dual-Camera Compressive Hyperspectral Imaging}

\author{%
  Ge Meng, 
  Zhongnan Cai,
  Jingyan Tu,
  Yingying Wang,
  Chenxin Li,
  Yue Huang, 
  Xinghao Ding\thanks{Corresponding Author.} \\
  \\
  Xiamen University\\
  \\
  mengge0001@gmail.com, wangyingying7@stu.xmu.edu.cn, yhuang2010@xmu.edu.cn
}

\begin{document}

\maketitle

\begin{abstract}
Panchromatic (PAN) -assisted Dual-Camera Compressive Hyperspectral Imaging (DCCHI) is a key technology in snapshot hyperspectral imaging. Existing research primarily focuses on exploring spectral information from 2D compressive measurements and spatial information from PAN images in an explicit manner, leading to a bottleneck in HSI reconstruction. Various physical factors, such as temperature, emissivity, and multiple reflections between objects, play a critical role in the process of a sensor acquiring hyperspectral thermal signals. Inspired by this, we attempt to investigate the interrelationships between physical properties to provide deeper theoretical insights for HSI reconstruction. In this paper, we propose a Physics-Informed Cross-Modal State Space Model Network (PCMamba) for DCCHI, which incorporates the forward physical imaging process of HSI into the linear complexity of Mamba to facilitate lightweight and high-quality HSI reconstruction. Specifically, we analyze the imaging process of hyperspectral thermal signals to enable the network to disentangle the three key physical properties—temperature, emissivity, and texture. By fully exploiting the potential information embedded in 2D measurements and PAN images, the HSIs are reconstructed through a physics-driven synthesis process. Furthermore, we design a Cross-Modal Scanning Mamba Block (CSMB) that introduces inter-modal pixel-wise interaction with positional inductive bias by cross-scanning the backbone features and PAN features. Extensive experiments conducted on both real and simulated datasets demonstrate that our method significantly outperforms SOTA methods in both quantitative and qualitative metrics. 

\end{abstract}

\section{Introduction}
   

Hyperspectral images (HSIs) have multiple continuous and narrow spectral bands, which can capture the reflective properties of objects in different bands and store richer information. Based on this property, HSIs have been widely applied in multiple fields, for example, medical imaging~\cite{liu2019flexible,meng2020snapshot}, remote sensing~\cite{yuan2017hyperspectral,shimoni2019hypersectral}, material classification~\cite{keshava2004distance,khan2018modern}, object tracking~\cite{uzkent2017aerial,li2022target}, etc.

Conventional hyperspectral imaging uses a single 1D or 2D sensor that captures HSIs by scanning spatial or spectral dimensions with long exposures, which limits its applicability to dynamic scenes. To address this, researchers have employed the coded aperture snapshot spectral imaging (CASSI) system to capture the 3D HSI cube~\cite{arce2013compressive,llull2013coded}. CASSI leverages the sparsity of spectral data to capture compressed 2D measurements by modulating spectral signals with coded apertures and dispersive elements. However, high-quality reconstruction demands multiple acquisitions of the same scene with varying coded apertures to enrich the available measurements. Dual Camera Compressed Hyperspectral Imaging (DCCHI)~\cite{wang2018high,xie2022dual}, based on CASSI with the addition of a beam splitter and a grayscale camera, produces a higher-quality HSI reconstruction than CASSI by fusing the compressed and PAN images while maintaining the advantages of snapshots.

\begin{wrapfigure}{r}{7cm}
 \vspace{-8pt}
    \includegraphics[width=\linewidth]{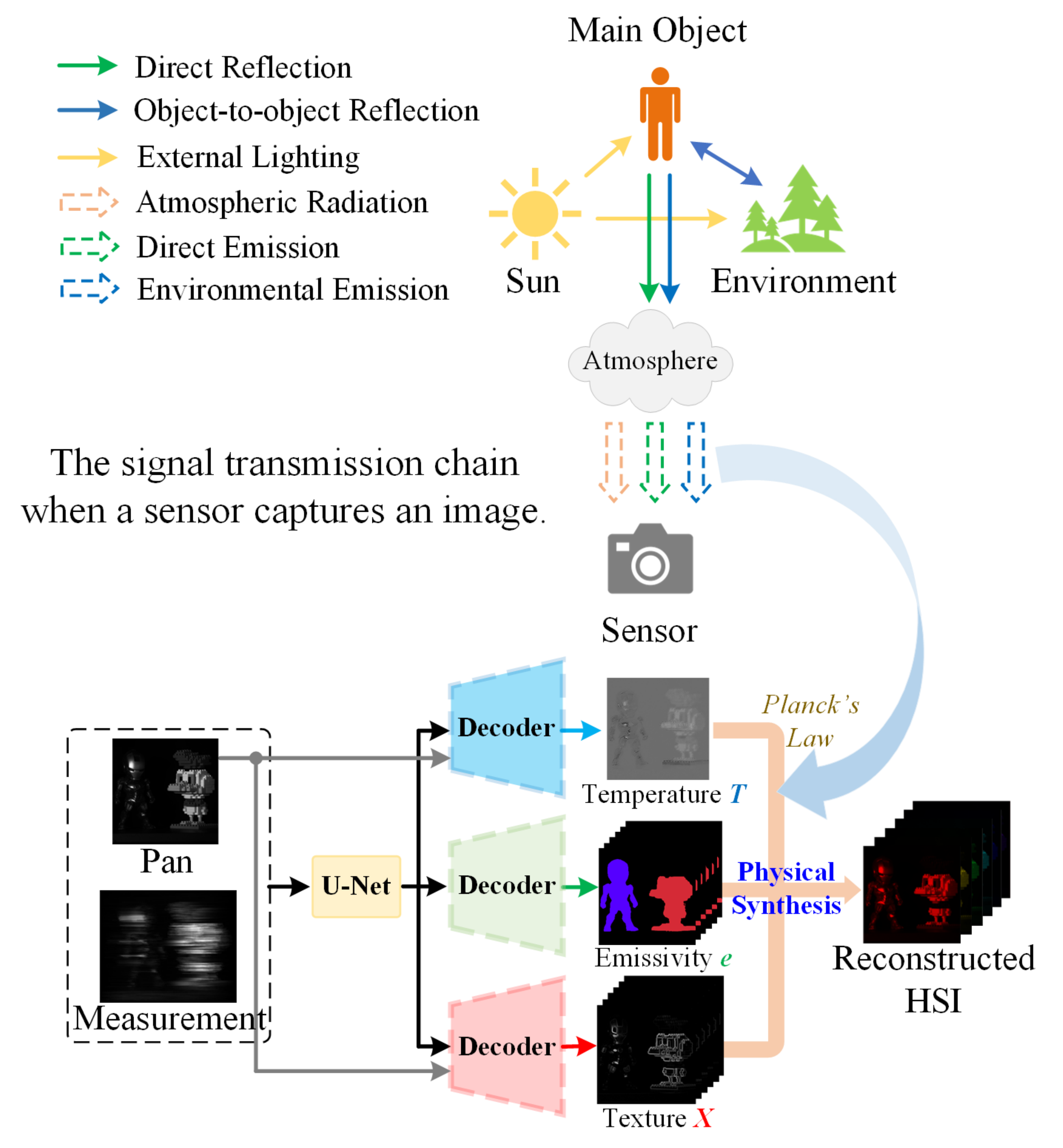}
    \caption{Our physics-informed HSI reconstruction method.}
    \label{fig:intro}
\end{wrapfigure}

Traditional methods utilize hand-designed priors for reconstruction, such as sparsity~\cite{lin2014spatial,wang2016simultaneous}, non-local similarity~\cite{fu2016exploiting,he2020non}, low-rank~\cite{liu2018rank}, and total variation~\cite{kittle2010multiframe,wang2015dual}. However, these methods require manual adjustment of the parameters, which often leads to mismatches between the prior assumptions and the actual problem. 
In deep learning-based methods, end-to-end methods~\cite{yuan2018hyperspectral,meng2020end} treat the HSI reconstruction process as a black box, learning the transformation that maps spatial and spectral information to reconstructed HSIs. Additionally, some deep unfolding methods~\cite{cai2022degradation,li2023pixel} attempt to incorporate physical priors into model training to enhance the model's interpretability. The prior modules in these methods typically require a denoiser for multi-stage optimization, which is designed to effectively utilize both spatial and spectral information. However, in the aforementioned methods, researchers primarily focus on explicitly learning spatial and spectral information, often overlooking the physical imaging process of HSI~\cite{bao2023heat}. This has led to a bottleneck in HSI reconstruction.

In recent years, vision Transformer (ViT)-based HSI reconstruction methods~\cite{cai2022mask,li2023pixel,yao2024specat} have emerged due to their capability to capture global dependencies and non-local similarities.
However, due to the significantly higher dimensionality of HSI data compared to traditional RGB images, the computational complexity of ViT grows quadratically with the increase in input sequence length. Visual Mamba (VMamba) models sequential dependencies by dividing an image into sequential blocks while maintaining linear computational complexity~\cite{liu2025vmamba}. 
This approach has been widely applied in various fields, including remote sensing image segmentation~\cite{zhu2024samba,cao2024remote}, multimodal detection~\cite{ren2024remotedet}, and image super-resolution~\cite{qiao2024hi,ren2024mambacsr}. However, VMamba assumes that the relationships between input sequences diminish as the sequential distance increases, which limits the ability to learn compact correlations between multimodal information. In addition, VMamba's pixel-wise scanning of high-dimensional data imposes a heavy computational burden, leading to significant resource consumption.

To address the issues mentioned above, we propose a physics-informed cross-modal state space model (SSM) network (PCMamba) for DCCHI. PCMamba integrates the physical imaging process of hyperspectral signals with the linear complexity advantage of Mamba, effectively enhancing the quality of HSI reconstruction while preserving a lightweight computational cost. Specifically, we first analyze the physical process involved when the sensor captures spectral signals, as shown in Fig. \ref{fig:intro}. The thermal signals produced by an object in a specific band stem from both its own direct thermal emission and the environmental emission of surrounding objects. The former is governed by Planck’s law, the object's temperature, and its emissivity, while the latter encapsulates the primary texture information in the HSI. Both signals are captured by the sensor after propagating through the atmosphere. Motivated by this observation, we aim to disentangle the object's temperature, emissivity, and texture, and subsequently reconstruct the HSI through a physical synthesis process. This physics-informed feature learning approach fully utilizes the latent information embedded in 2D compressed measurements and PAN images, moving beyond the sole reliance on explicit spectral and spatial information. To achieve this goal, we design a Cross-Modal Scanning Mamba Block (CSMB) that performs inter-modal cross-scanning between the backbone features and PAN features, without pixel position repetition. This scanning scheme reduces the sequence length by half, while encouraging the model to learn more compact pixel-wise inter-modal interactions with a positional inductive bias. 

Our contributions can be summarized as follows:
\begin{itemize}
    \item [1)] We propose PCMamba, a physics-informed cross-modal SSM network for DCCHI. PCMamba integrates the linear complexity of Mamba with the forward physical imaging process of HSI, enabling both lightweight and high-quality HSI reconstruction.
    \item [2)] This is the first attempt to tackle the DCCHI task from the perspective of HSI thermal signal imaging, aiming to overcome the existing bottleneck in HSI reconstruction by exploring the latent physical properties embedded in the input image.
    \item [3)] We design a Cross-Modal Scanning Mamba Block (CSMB) that performs cross-scanning of features from different modalities without overlapping pixel positions. This design shortens the sequence length processed by the model and enhances its ability to learn more compact inter-modal associations.
    \item [4)] Experiments on both real and simulated datasets demonstrate the effectiveness and efficiency of the proposed method. Ablation studies further validate the contribution of each module.
\end{itemize}

\section{Related Work}
\label{sec:related work}


\subsection{DCCHI System}

\begin{wrapfigure}{r}{6cm}
 \vspace{-8pt}
    \includegraphics[width=\linewidth]{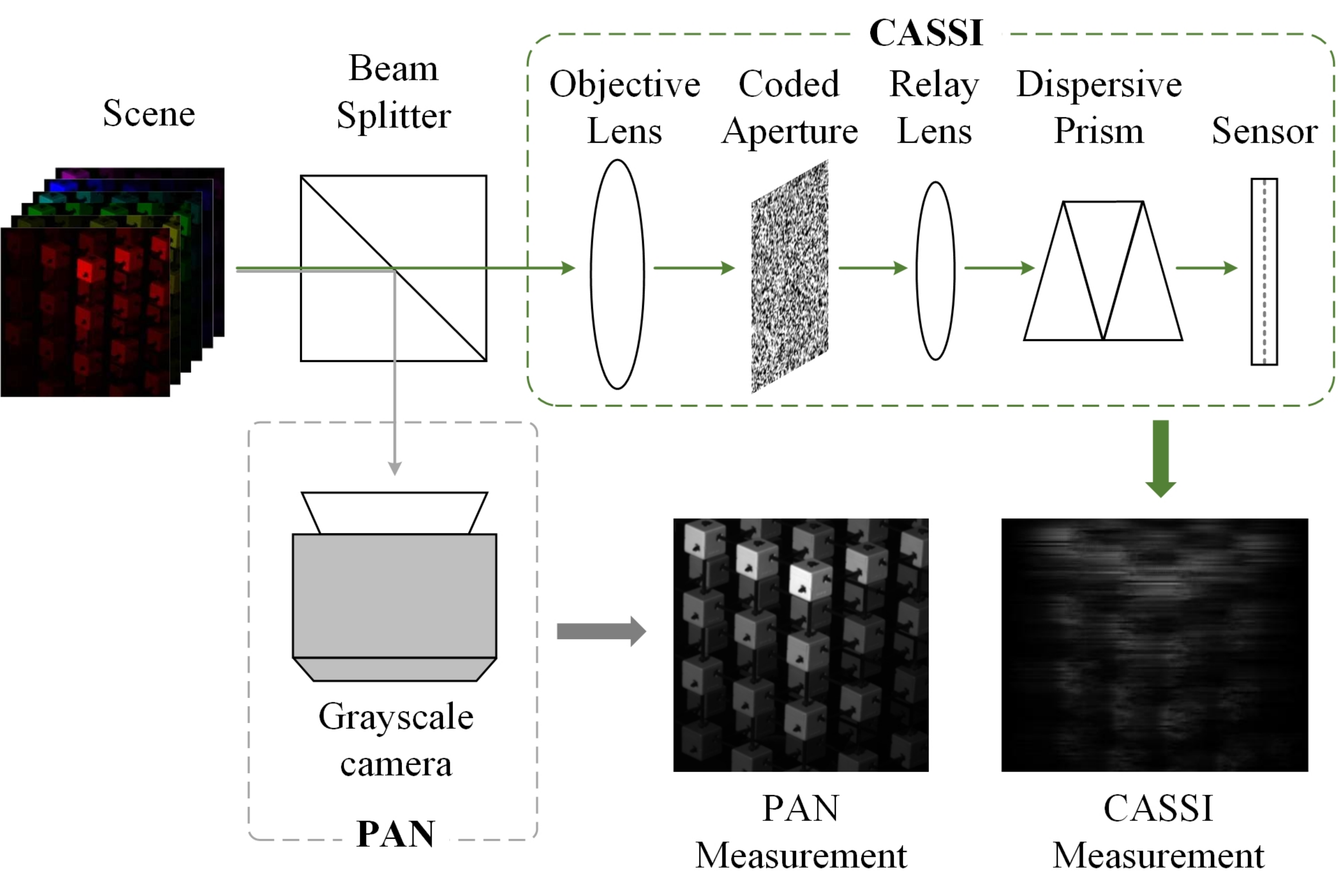}
    \caption{The dual-camera compressive hyperspectral imaging
system.}
    \label{fig:DCHHI}
\end{wrapfigure}

The principle of DCCHI is shown in Fig. \ref{fig:DCHHI}, which consists of a beam splitter, a PAN camera branch and a CASSI branch. The beam splitter splits the incident light equally in two directions, one direction is captured by the CASSI branch, which compresses the 3D HSI cube into a measurement by spatial and spectral modulation, and the other direction is captured by the PAN camera branch to generate a grayscale measurement.

The measurement $\mathbf{y^{c}}$ obtained by CASSI branch modulation can be formulated as

\begin{equation}
  \mathbf{y^{c}} =\mathbf{\Phi^{c}}\mathbf{x} + \mathbf{g^{c}},
  \label{eq:measurementyc}
\end{equation}
where $\mathbf{x}$ is the 3D HSI cube of the target scene, $\Phi^{c}$ is the coded aperture, and $\mathbf{g^{c}}$ is the Gaussian noise generated during the imaging process.
Likewise, the imaging model of the PAN camera branch can be described as
\begin{equation}
  \mathbf{y^{P}} =\mathbf{\Phi^{P}}\mathbf{x} + \mathbf{g^{P}},
  \label{eq:measurementyp}
\end{equation}
To faciliate the subsequent discussion of the imaging model, we define $\mathbf{y}$, $\mathbf{\Phi}$, $\mathbf{g}$ as
\begin{equation}
  \mathbf{y}=\begin{bmatrix}
    \mathbf{y^c}\\\mathbf{y^P}
    \end{bmatrix},
    \mathbf{\Phi}=\begin{bmatrix}
    \mathbf{\Phi^c}\\\mathbf{\Phi^P}
    \end{bmatrix},
    \mathbf{g}=\begin{bmatrix}
    \mathbf{g^c}\\\mathbf{g^P}
    \end{bmatrix},
  \label{eq:measurementyn}
\end{equation}
Based on the above definition, the imaging model of DCCHI can be expressed as
\begin{equation}
  \mathbf{y} =\mathbf{\Phi}\mathbf{x} + \mathbf{g}.
  \label{eq:measurementy}
\end{equation}


\subsection{Traditional HSI reconstruction methods}
Traditional HSI reconstruction methods are mainly relied on hand-crafted priors~\cite{yuan2016generalized,liu2018rank,he2020non}. For example, GAP-TV~\cite{yuan2016generalized} introduces a generalized alternating projection algorithm using total variation minimization. Twist~\cite{bioucas2007new} proposes a two-step iterative shrinkage / thresholding algorithm for reconstructing missing samples. Non-local similarity and low-rank regularization~\cite{fu2016exploiting,liu2018rank,he2020non} are applied to explore spatial and spectral correlations. Sparse representation~\cite{lin2014spatial,arguello2013higher,wang2016simultaneous} models image sparsity by learning complete dictionaries. In ~\cite{yang2014compressive}, the image is reconstructed by learning a Gaussian mixture model of the signal. However, these model-driven methods lack efficiency and flexibility due to the need for manual tuning of extensive parameters. 

\subsection{Deep Learning-based DCCHI Methods}
Using PAN to assist HSI reconstruction can effectively address the inherent limitations of CASSI. PFusion~\cite{he2021fast} uses RGB measurements to estimate the spatial coefficients and employs CASSI measurements to provide the spectral basis, thereby exploring the low-dimensional spectral subspace property of the HSI, which consists of spectral basis and spatial coefficients. PIDS~\cite{chen2023prior} utilizes the RGB image as a prior image to provide valuable semantic information. In2SET~\cite{Wang_2024_CVPR} employs a novel attention mechanism to capture both intra-similarity and inter-similarity between spectral and PAN images. Besides, some ViT-based methods attempt to recover HSI from a single measurement. MST~\cite{cai2022mask} utilizes transformer to capture the inter-spectral similarity of HSI. DAUHST~\cite{cai2022degradation} customizes a Half-Shuffle Transformer to capture both local content and non-local dependencies. PADUT~\cite{li2023pixel} proposes the nonlocal spectral transformer for modeling spatial and spectral similarity at a fine-grained level. SPECAT~\cite{yao2024specat} uses Swintransformer~\cite{liu2021swin} to model spatial sparsity. These methods show superior performance on HSI reconstruction than previous methods. However, these methods focus on explicitly exploring spatial and spectral information, while neglecting the underlying physical properties of HSI.

\section{Method}
\label{sec:method}
\begin{figure*}[t]
  \centering
   \includegraphics[width=1\linewidth]{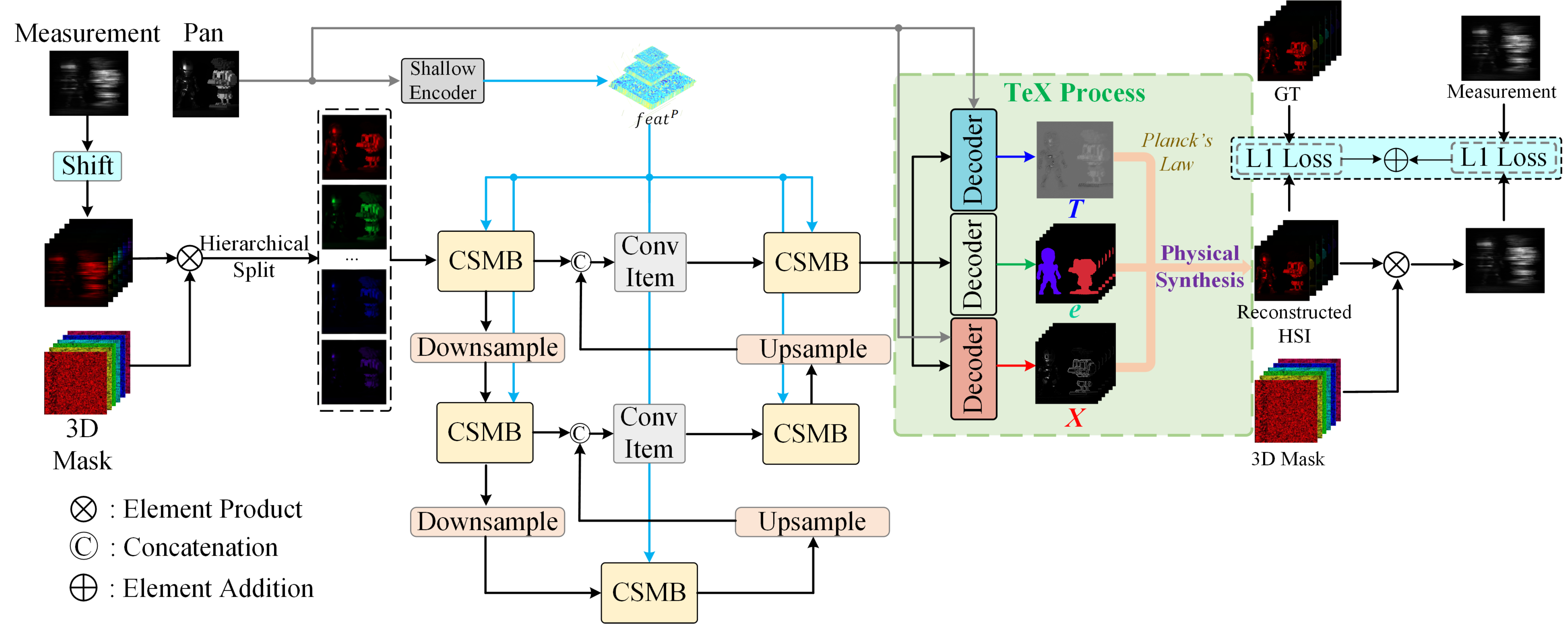}
   \caption{Overview of PCMamba. PCMamba consists of a state space model network with a U-net architecture composed of Cross-Modal Scanning Mamba Blocks (CSMB), and the physical synthesis process of the HSI, which combines Planck's law and TeX decomposition.}
   \label{fig:pcmamba}
\end{figure*}

Fig. \ref{fig:pcmamba} illustrates the overall architecture of our proposed PCMamba, which consists of two parts: a state space model (SSM) network with a U-net architecture composed of Cross-Modal Scanning Mamba Blocks (CSMBs), and the physical synthesis process of the HSI. The details are illustrated as below.

\subsection{TeX decomposition}
During the process of capturing hyperspectral signals, the sensor mixes the temperature (\emph{T}, physical status), emissivity (\emph{e}, material fingerprint), and texture of the object (\emph{X}, surface geometry) in the photon flux~\cite{bao2023heat}.
These physical properties collectively contribute to both the direct emission of the primary target and the environmental emission from surrounding objects.
\begin{equation}
    \begin {aligned}
    \mathcal{S} _{\lambda }^{\alpha } =\Phi _{\lambda }^{\alpha }+\Psi _{\lambda }^{\alpha },
    \end{aligned}
\end{equation}
where $\Phi _{\lambda }^{\alpha }$ is the direct emission from the object $\alpha$ at wavelength $\lambda$, and $\Psi _{\lambda }^{\alpha }$ is environmental emission. Furthermore, the direct emission $\Phi _{\lambda }^{\alpha }$ consists of two components: the blackbody radiation of the object $\alpha$, governed by Planck's law, and its emissivity.
\begin{equation}
    \begin {aligned}
    \Phi _{\lambda }^{\alpha } = \mathit{e}_{\lambda }^{\alpha }  B_{\lambda } \left ( T \right ),
    \end{aligned}
\end{equation}
\begin{equation}
    \begin {aligned}
    B_{\lambda } \left ( T \right ) =\frac{2\pi hc^{2} }{\lambda ^{5} } \frac{1}{e^{\frac{hc}{\lambda kT} } -1},
    \end{aligned}
    \label {eq:black}
\end{equation}
where $\emph{e}_{\lambda }^{\alpha }$ represents the emissivity of the object $\alpha$ at wavelength $\lambda$, and $B_{\lambda } \left ( T \right )$ represents its blackbody radiation. Eq. (\ref{eq:black}) is Planck's law, where $h$ is Planck constant, $k$ is Boltzmann constant, and $c$ is the speed of light. It is evident that the blackbody radiation of an object is solely determined by its wavelength $\lambda$ and temperature \emph{T}.

\begin{figure*}[t]
  \centering
   \includegraphics[width=1\linewidth]{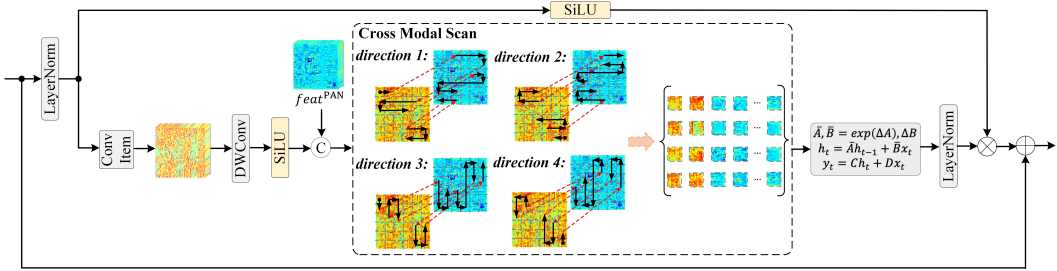}
   \caption{Illustration of Cross-Modal Scanning Mamba Block (CSMB). CSMB performs non-overlapping cross-scanning between backbone features and PAN features to learn more compact inter-modal correlations.}
   \label{fig:csmb}
\end{figure*}

An important fact is that the surface texture of the object is obscured by its direct thermal emission,  a phenomenon known as the ``ghosting effect''~\cite{gurton2014enhanced}. The structural information observed in the image primarily comes from external light sources and environmental emissions, as shown in Fig. \ref{fig:intro}. Therefore, to fully describe the hyperspectral signal, it is necessary to consider the emission contributed by the environment. Given the emissivity $\emph{e}_{\lambda }^{\alpha }$ of object $\alpha$ at wavelength $\lambda$, its corresponding environmental emission $\Psi _{\lambda }^{\alpha }$ can be calculated as
\begin{equation}
    \begin {aligned}
    \Psi _{\lambda }^{\alpha } =\left ( 1- e _{\lambda }^{\alpha } \right ) X _{\lambda }^{\alpha },
    \end{aligned}
\end{equation}
\begin{equation}
    \begin {aligned}
    X _{\lambda }^{\alpha } = \sum_{i=1}^{N-1}V_{\alpha \beta _{i} }\mathcal{S} _{\lambda }^{\beta _{i}  } ,
    \end{aligned}
\end{equation}
where $\beta _{i}$ is the $i$-th object surrounding $\alpha$, and $V_{\alpha \beta _{i}}$ represents the linear combination vector between $\beta _{i}$ and $\alpha$. $\Psi$ represents multiple reflections between objects and contains the main texture information.

Furthermore, the hyperspectral signals, after passing through the atmosphere, are transmitted to the sensor along with atmospheric radiation
\begin{equation}
    \begin {aligned}
    \mathcal{T} =\gamma \mathcal{S} +\left ( 1-\gamma  \right ) \Lambda ,
    \end{aligned}
\end{equation}
where $\gamma$ represents the transmissivity of atmosphere~\cite{Incropera_DeWitt_Bergman_Lavine_2018}, $\Lambda$ is atmospheric radiation, and $\mathcal{T}$ represents the total radiation signal.

Finally, the hyperspectral signal captured by the sensor can be rewritten as
\begin{equation}
    \begin {aligned}
    \mathcal{T} =\gamma \left ( \mathit{e}_{\lambda }  B_{\lambda } \left ( T \right ) + \left ( 1- e _{\lambda } \right ) X _{\lambda } \right ) + \left ( 1-\gamma  \right ) \Lambda ,
    \end{aligned}
\end{equation}
Due to the absorption of spectral signals by water vapor and carbon dioxide in the atmosphere, $\gamma$ is typically close to 1, resulting in the approximation of $\mathcal{T}$ as
\begin{equation}
    \begin {aligned}
    \mathcal{T} =\mathit{e}_{\lambda }  B_{\lambda } \left ( T \right ) + \left ( 1- e _{\lambda } \right ) X _{\lambda }.
    \end{aligned}
\end{equation}
This implies that the HSI can be synthesized through a forward physical process if the temperature $T$, emissivity $e$, and texture $X$ can be accurately obtained.

\subsection{Hyperspectral Signals}
For HSIs, the spectral signals captured at wavelength $\lambda$ is equal to the superimposition of signals over the entire wavelength range [$\lambda_{min}$, $\lambda_{max}$]
\begin{equation}
    \begin{aligned}
    \int_{\lambda_{min}  }^{\lambda_{max}} \mathcal{S} _{\lambda }^{\alpha} d\lambda &= \int_{\lambda_{min}  }^{\lambda_{max}}e_{\lambda }^{\alpha} B_{\lambda }^{\alpha} \left ( T \right ) d\lambda \\
    &+ \int_{\lambda_{min}  }^{\lambda_{max}}\left ( 1-e_{\lambda }^{\alpha} \right ) \sum_{i=1}^{N-1}V_{\alpha \beta_{i} } \mathcal{S} _{\lambda }^{\beta_{i}} d\lambda,
    \end{aligned}
\end{equation}
Available data in~\cite{bao2023heat} indicates that, the emissivity $e_{\lambda}$ of most objects remains relatively constant across the short working wavelengths. Therefore, we assume a constant emissivity for a specific material
\begin{equation}
    \begin{aligned}
    \int_{\lambda_{min}  }^{\lambda_{max}} \mathcal{S} _{\lambda }^{\alpha} d\lambda &= e_{\lambda }^{\alpha} \int_{\lambda_{min}  }^{\lambda_{max}} B_{\lambda }^{\alpha} \left ( T \right ) d\lambda \\
    &+ \left ( 1-e_{\lambda }^{\alpha} \right ) \int_{\lambda_{min}  }^{\lambda_{max}} \sum_{i=1}^{N-1}V_{\alpha \beta_{i} } \mathcal{S} _{\lambda }^{\beta_{i}} d\lambda.
    \end{aligned}
\end{equation}

\subsection{Cross-Modal Scanning Mamba Block (CSMB)}
SSM utilizes a framework of linear ordinary differential equations to map inputs to outputs through hidden state. For a system with input excitation $x(t) \in \mathbb{R}^L$, hidden state $h(t) \in \mathbb{C}^N$ and output response $y(t) \in \mathbb{R}^L$, the model can be formulated as
\begin{equation}
\begin {aligned}
    &h'(t) = \mathbf{A}h(t) + \mathbf{B}x(t),\\
    &y(t) = \mathbf{C}h(t) + \mathbf{D}x(t),
\label{eq:ssm1}
\end{aligned}
\end{equation}
where $\mathbf{A} \in \mathbb{C}^{N \times N}$, $\mathbf{B}, \mathbf{C} \in \mathbb{C}^{N}$ and $\mathbf{D} \in \mathbb{C}^{1}$ are weighting parameters. Subsequently, a discretization of Eq. (\ref{eq:ssm1}) is usually obtained using a zero-order keeper (ZOH)
\begin{equation}
    \begin{array}{c}
    \overline{\mathbf{A}}=\exp (\Delta \mathbf{A}), \\
    \overline{\mathbf{B}}=(\Delta \mathbf{A})^{-1}(\exp (\Delta \mathbf{A})-I) \cdot \Delta \mathbf{B},
    \end{array}
\end{equation}
where $\Delta$ is a time scale parameter used to transform the continuous parameters $\mathbf{A}$, $\mathbf{B}$ into discrete parameters $\overline{\mathbf{A}}$, $\overline{\mathbf{B}}$. The discretized Eq. (\ref{eq:ssm1}) can be written as
\begin{equation}
    \begin {aligned}
    &h_{t}=\overline{\mathbf{A}} h_{t-1}+\overline{\mathbf{B}} x_{t}, \\
    &y_{t}=\mathbf{C} h_{t}+\mathbf{D} x_{t},
    \end{aligned}
\end{equation}
Given that the dimensionality of HSI data is several orders of magnitude higher than that of conventional RGB image, we adopt the SSM with linear complexity to keep the network lightweight. We aim to facilitate effective interaction between information from compressed measurements and PAN images, while mitigating interference from redundant spectral information. We first employ a shallow encoder to extract multi-scale PAN features
\begin{equation}
    \begin {aligned}
    \left [ feat_{1}^{P}, feat_{2}^{P},\cdots ,feat_{M}^{P} \right ] =\mathrm{SE}  \left ( P \right ),
    \end{aligned}
\end{equation}
where $P$ denotes a PAN image, SE($\cdot$) is a shallow encoder, and the subscript $M$ denotes the PAN features at different scales. Similarly, the backbone features are represented as $feat_{N}^{B}$. Then we ensure that $feat_{k}^{P}$ and $feat_{k}^{B}$ have the same spatial size and perform cross-scanning on them without pixel position overlap, as shown in Fig. \ref{fig:csmb}
\begin{equation}
    \begin {aligned}
    feat^\mathrm{B}_{k+1} &= \mathrm{CS} \left( \mathrm{LN} \left( feat^B_k \right), feat^P_k \right) \odot \\
    &\mathrm{SiLU} \left( \mathrm{LN} \left( feat^B_k \right) \right) + feat^\mathrm{B}_{k}.
    \end{aligned}
\end{equation}
where $\operatorname{CS\left( \cdot \right) }$ represents the cross-modal scanning operation which employs the following operation sequence: $DWConv \to SiLU \to SSM \to LN$. This non-overlapping pixel position scanning method helps to suppress redundant spectral information between bands, encourages the network to learn a more compact pixel-wise inductive bias between emissivity and texture, and simultaneously reduces the sequence length processed by the SSM by half. 

Finally, three decoders are applied to the output features of the U-net to generate the desired temperature \emph{T}, emissivity \emph{e}, and texture \emph{X}. Since the PAN image theoretically shares the same temperature properties as the measurement, we leverage it to facilitate the generation of \emph{T}. Likewise, the PAN is utilized to inject additional structural details into the generation of \emph{X}, as shown in Fig. \ref{fig:pcmamba}.

\subsection{Loss Function}
In this paper, we use L1 loss to optimize the reconstructed HSI at the pixel level
\begin{equation}
\begin{split}
\mathcal{L}_{Rec} = \frac{1}{H \times W \times C} \sum_{i=0} \left | \widehat{Y}\left ( i \right ) - Y \left ( i \right )   \right |,
\end{split}
\end{equation}
where $C$ is the number of channels in the reconstructed HSI, $\widehat{Y}\left ( i \right )$ represents the predicted value for pixel $i$, and $Y \left ( i \right )$ is its corresponding ground truth. 
Besides, we ensure the rationality of the entire reconstruction process by modulating the reconstruction result of PCMamba to generate a 2D measurement consistent with the network input
\begin{equation}
\begin{gathered}
\mathcal{L} _{\mathcal{M} } ={\left \| \mathcal{M}\odot \widehat{Y} - \mathcal{X} \right \|}^{2}_{1} ,
\end{gathered}
\end{equation}
where $\mathcal{M}$ is the modulated mask and $\mathcal{X}$ is the input 2D measurement. The total loss $\mathcal{L}_{total}$ is defined by combing the reconstruction loss $\mathcal{L}_{Rec}$ and the imaging process consistency loss $\mathcal{L} _{\mathcal{M} }$
\begin{equation}
\begin{gathered}
\mathcal{L} _{total} = \mathcal{L}_{Rec} + \mathcal{L} _{\mathcal{M} }.
\end{gathered}
\end{equation}

\begin{table*}[h]
\Huge
\renewcommand{\arraystretch}{3} 
\centering
\caption{Comparison of different methods on 10 simulated scenes including FLOPS, PSNR (upper entry) and SSIM (lower entry). ``Method-RGB'' denotes the use of RGB observations, and ``Method-PAN'' denotes grayscale observations in the PAN image. Notably, ``Method-PAN'' is modified from other HSI reconstruction methods in the DCCHI reconstruction task. The best values are bolded.} 
\normalsize
\resizebox{\linewidth}{!}{
\begin{tabular}{l c ccccccccccc}

\hline
\LARGE\textbf{Methods} &  \LARGE\textbf{GFLOPs} & \LARGE\textbf{Scene1} & 
\LARGE\textbf{Scene2} & \LARGE\textbf{Scene3} & \LARGE\textbf{Scene4} & \LARGE\textbf{Scene5} & \LARGE\textbf{Scene6} &\LARGE\textbf{Scene7} 
& \LARGE\textbf{Scene8} 
&\LARGE \textbf{Scene9} 
& \LARGE\textbf{Scene10} 
& \LARGE\textbf{Avg} \\
\hline
\LARGE\textbf{PFsion-RGB~\cite{he2021fast}} & \textbf{\LARGE-} & \makecell{\LARGE40.09 \\ \LARGE0.979} & \makecell{\LARGE38.84\\\LARGE0.968} & \makecell{\LARGE38.70\\\LARGE0.966} & \makecell{\LARGE46.65\\\LARGE0.936} & \makecell{\LARGE32.07\\\LARGE0.980} & \makecell{\LARGE37.12\\\LARGE0.980} & \makecell{\LARGE39.74\\\LARGE0.964} & \makecell{\LARGE36.75\\\LARGE0.965} & \makecell{\LARGE34.52\\\LARGE0.931} & \makecell{\LARGE35.53\\\LARGE0.979} & \makecell{\LARGE38.00\\\LARGE0.965} \\

\hline
\LARGE\textbf{PIDS-RGB~\cite{chen2023prior}} & \textbf{\LARGE-} & \makecell{\LARGE42.09 \\ \LARGE0.983} & \makecell{\LARGE40.08\\\LARGE0.949} & \makecell{\LARGE41.50\\\LARGE0.968} & \makecell{\LARGE48.55\\\LARGE0.989} & \makecell{\LARGE40.05\\\LARGE0.982} & \makecell{\LARGE39.00\\\LARGE0.974} & \makecell{\LARGE36.63\\\LARGE0.940} & \makecell{\LARGE37.02\\\LARGE0.948} & \makecell{\LARGE38.82\\\LARGE0.953} & \makecell{\LARGE38.64\\\LARGE0.980} & \makecell{\LARGE40.24\\\LARGE0.967} \\
\hline
\LARGE\textbf{TV-PAN~\cite{wang2015dual}} & \textbf{\LARGE-} & \makecell{\LARGE35.81 \\ \LARGE0.947} & \makecell{\LARGE33.22\\\LARGE0.885} & \makecell{\LARGE31.07\\\LARGE0.879} & \makecell{\LARGE40.11\\\LARGE0.947} & \makecell{\LARGE33.32\\\LARGE0.944} & \makecell{\LARGE34.62\\\LARGE0.943} & \makecell{\LARGE31.09\\\LARGE0.885} & \makecell{\LARGE32.31\\\LARGE0.916} & \makecell{\LARGE29.36\\\LARGE0.862} & \makecell{\LARGE33.84\\\LARGE0.953} & \makecell{\LARGE33.47\\\LARGE0.910} \\
\hline
\LARGE\textbf{PIDS-PAN~\cite{chen2023prior}} & \textbf{\LARGE-} & \makecell{\LARGE39.82 \\ \LARGE0.977} & \makecell{\LARGE37.07\\\LARGE0.921} & \makecell{\LARGE37.72\\\LARGE0.950} & \makecell{\LARGE46.78\\\LARGE0.978} & \makecell{\LARGE37.45\\\LARGE0.973} & \makecell{\LARGE37.74\\\LARGE0.963} & \makecell{\LARGE32.90\\\LARGE0.896} & \makecell{\LARGE31.66\\\LARGE0.915} & \makecell{\LARGE34.35\\\LARGE0.902} & \makecell{\LARGE38.58\\\LARGE0.972} & \makecell{\LARGE37.41\\\LARGE0.945} \\
\hline
\LARGE\textbf{BiSRNet-PAN~\cite{cai2024binarized}} & \textbf{\LARGE1.33} & \makecell{\LARGE35.02 \\ \LARGE0.945} & \makecell{\LARGE34.13\\\LARGE0.914} & \makecell{\LARGE31.50\\\LARGE0.883} & \makecell{\LARGE35.88\\\LARGE0.895} & \makecell{\LARGE33.70\\\LARGE0.935} & \makecell{\LARGE35.58\\\LARGE0.925} & \makecell{\LARGE32.31\\\LARGE0.900} & \makecell{\LARGE32.73\\\LARGE0.903} & \makecell{\LARGE31.37\\\LARGE0.899} & \makecell{\LARGE34.48\\\LARGE0.936} & \makecell{\LARGE33.67\\\LARGE0.914} \\
\hline
\LARGE\textbf{CST-PAN~\cite{cai2022coarse}} & \LARGE25.40 & \makecell{\LARGE37.44 \\ \LARGE0.975} & \makecell{\LARGE38.91\\\LARGE0.978} & \makecell{\LARGE36.79\\\LARGE0.969} & \makecell{\LARGE42.27\\\LARGE0.983} & \makecell{\LARGE36.57\\\LARGE0.982} & \makecell{\LARGE38.91\\\LARGE0.984} & \makecell{\LARGE36.87\\\LARGE0.967} & \makecell{\LARGE35.91\\\LARGE0.980} & \makecell{\LARGE35.87\\\LARGE0.973} & \makecell{\LARGE37.93\\\LARGE0.990} & \makecell{\LARGE37.75\\\LARGE0.978} \\
\hline
\LARGE\textbf{HDNet-PAN~\cite{hu2022hdnet}} & \LARGE144.31 & \makecell{\LARGE38.06 \\ \LARGE0.976} & \makecell{\LARGE39.79\\\LARGE0.982} & \makecell{\LARGE38.21\\\LARGE0.973} & \makecell{\LARGE42.79\\\LARGE0.983} & \makecell{\LARGE37.22\\\LARGE0.983} & \makecell{\LARGE39.26\\\LARGE0.986} & \makecell{\LARGE37.41\\\LARGE0.968} & \makecell{\LARGE36.51\\\LARGE0.982} & \makecell{\LARGE36.64\\\LARGE0.976} & \makecell{\LARGE37.52\\\LARGE0.988} & \makecell{\LARGE38.34\\\LARGE0.980} \\
\hline
\LARGE\textbf{MST++-PAN~\cite{cai2022mst++}} & \LARGE17.69 & \makecell{\LARGE38.38 \\ \LARGE0.977} & \makecell{\LARGE40.47\\\LARGE0.980} & \makecell{\LARGE37.70\\\LARGE0.968} & \makecell{\LARGE43.88\\\LARGE0.984} & \makecell{\LARGE37.75\\\LARGE0.983} & \makecell{\LARGE39.42\\\LARGE0.985} & \makecell{\LARGE37.48\\\LARGE0.964} & \makecell{\LARGE37.38\\\LARGE0.982} & \makecell{\LARGE38.82\\\LARGE0.980} & \makecell{\LARGE39.04\\\LARGE0.989} & \makecell{\LARGE39.03\\\LARGE0.980} \\
\hline
\LARGE\textbf{DAUHST-PAN-2stg~\cite{cai2022degradation}} & \LARGE16.79 & \makecell{\LARGE40.78 \\ \LARGE0.983} & \makecell{\LARGE43.13\\\LARGE0.987} & \makecell{\LARGE41.73\\\LARGE0.980} & \makecell{\LARGE47.09\\\LARGE0.990} & \makecell{\LARGE39.84\\\LARGE0.987} & \makecell{\LARGE40.90\\\LARGE0.986} & \makecell{\LARGE39.75\\\LARGE0.976} & \makecell{\LARGE38.98\\\LARGE0.981} & \makecell{\LARGE41.29\\\LARGE0.983} & \makecell{\LARGE40.04\\\LARGE0.988} & \makecell{\LARGE40.22\\\LARGE0.983} \\
\hline
\LARGE\textbf{DAUHST-PAN-3stg~\cite{cai2022degradation}} & \LARGE24.70 & \makecell{\LARGE40.22 \\ \LARGE0.983} & \makecell{\LARGE43.52\\\LARGE0.989} & \makecell{\LARGE41.74\\\LARGE0.981} & \makecell{\LARGE47.07\\\LARGE0.993} & \makecell{\LARGE38.81\\\LARGE0.985} & \makecell{\LARGE40.16\\\LARGE0.987} & \makecell{\LARGE39.86\\\LARGE0.978} & \makecell{\LARGE38.21\\\LARGE0.981} & \makecell{\LARGE40.63\\\LARGE0.983} & \makecell{\LARGE39.32\\\LARGE0.990} & \makecell{\LARGE40.95\\\LARGE0.985} \\
\hline
\LARGE\textbf{DAUHST-PAN-5stg~\cite{cai2022degradation}} & \LARGE40.51 & \makecell{\LARGE40.74 \\ \LARGE0.984} & \makecell{\LARGE44.00\\\LARGE0.989} & \makecell{\LARGE41.58\\\LARGE0.981} & \makecell{\LARGE46.84\\\LARGE0.991} & \makecell{\LARGE39.66\\\LARGE0.986} & \makecell{\LARGE40.89\\\LARGE0.987} & \makecell{\LARGE40.21\\\LARGE0.979} & \makecell{\LARGE38.72\\\LARGE0.983} & \makecell{\LARGE39.98\\\LARGE0.982} & \makecell{\LARGE40.10\\\LARGE0.989} & \makecell{\LARGE41.27\\\LARGE0.916} \\
\hline
\LARGE\textbf{DAUHST-PAN-9stg~\cite{cai2022degradation}} & \LARGE72.11 & \makecell{\LARGE41.59 \\ \LARGE0.985} & \makecell{\LARGE45.19\\\LARGE0.991} & \makecell{\LARGE43.47\\\LARGE0.984} & \makecell{\LARGE48.92\\\LARGE0.993} & \makecell{\LARGE40.27\\\LARGE0.988} & \makecell{\LARGE41.17\\\LARGE0.988} & \makecell{\LARGE40.73\\\LARGE0.979} & \makecell{\LARGE40.11\\\LARGE0.986} & \makecell{\LARGE43.50\\\LARGE0.988} & \makecell{\LARGE41.33\\\LARGE0.990} & \makecell{\LARGE42.62\\\LARGE0.987} \\
\hline
\LARGE\textbf{In2SET-2stg~\cite{Wang_2024_CVPR}} & \LARGE14.35 & \makecell{\LARGE40.33 \\ \LARGE0.983} & \makecell{\LARGE42.30\\\LARGE0.985} & \makecell{\LARGE40.34\\\LARGE0.977} & \makecell{\LARGE47.24\\\LARGE0.991} & \makecell{\LARGE39.42\\\LARGE0.986} & \makecell{\LARGE40.61\\\LARGE0.986}& \makecell{\LARGE39.46\\\LARGE0.975} & \makecell{\LARGE38.42\\\LARGE0.979} & \makecell{\LARGE40.37\\\LARGE0.981} & \makecell{\LARGE39.96\\\LARGE0.988} & \makecell{\LARGE40.84\\\LARGE0.983}  \\
\hline
\LARGE\textbf{In2SET-3stg~\cite{Wang_2024_CVPR}} & \LARGE20.79 & \makecell{\LARGE40.78 \\ \LARGE0.983} & \makecell{\LARGE43.13\\\LARGE0.987} & \makecell{\LARGE41.73\\\LARGE0.980} & \makecell{\LARGE47.09\\\LARGE0.990} & \makecell{\LARGE39.84\\\LARGE0.987} & \makecell{\LARGE40.90\\\LARGE0.986} & \makecell{\LARGE39.75\\\LARGE0.976} & \makecell{\LARGE38.98\\\LARGE0.981} & \makecell{\LARGE41.29\\\LARGE0.983} & \makecell{\LARGE40.04\\\LARGE0.988} & \makecell{\LARGE41.35\\\LARGE0.984} \\
\hline
\LARGE\textbf{In2SET-5stg~\cite{Wang_2024_CVPR}} & \LARGE33.66 & \makecell{\LARGE41.13 \\ \LARGE0.985} & \makecell{\LARGE44.43\\\LARGE0.990} & \makecell{\LARGE42.74\\\LARGE0.983} & \makecell{\LARGE47.29\\\LARGE0.993} & \makecell{\LARGE40.33\\\LARGE0.988} & \makecell{\LARGE40.95\\\LARGE0.987} & \makecell{\LARGE40.49\\\LARGE0.979} & \makecell{\LARGE39.15\\\LARGE0.982} & \makecell{\LARGE42.07\\\LARGE0.985} & \makecell{\LARGE39.44\\\LARGE0.987} & \makecell{\LARGE41.80\\\LARGE0.985} \\
\hline
\LARGE\textbf{In2SET-9stg~\cite{Wang_2024_CVPR}} & \LARGE59.40 & \makecell{\LARGE42.56 \\ \LARGE0.989} & \makecell{\LARGE46.42\\\LARGE0.994} & \makecell{\LARGE44.55\\\LARGE0.986} & \makecell{\textbf{\LARGE50.63}\\\textbf{\LARGE0.996}} & \makecell{\textbf{\LARGE42.01}\\\LARGE0.992} & \makecell{\LARGE42.49\\\LARGE0.991} & \makecell{\LARGE41.59\\\LARGE0.983} & \makecell{\LARGE40.53\\\LARGE0.989} & \makecell{\LARGE43.83\\\LARGE0.990} & \makecell{\LARGE42.33\\\LARGE0.994} & \makecell{\LARGE43.69\\\LARGE0.990} \\
\hline
\LARGE\textbf{SPECAT-PAN~\cite{yao2024specat}} & \LARGE12.40 & \makecell{\LARGE43.02 \\ \LARGE0.991} & \makecell{\LARGE45.79\\\LARGE0.994} & \makecell{\LARGE44.04\\\LARGE0.985} & \makecell{\LARGE47.09\\\LARGE0.993} & \makecell{\LARGE41.56\\\LARGE0.993} & \makecell{\LARGE44.32\\\LARGE0.994} & \makecell{\LARGE42.25\\\LARGE0.986} & \makecell{\LARGE43.15\\\LARGE0.993} & \makecell{\LARGE43.65\\\LARGE0.991} & \makecell{\textbf{\LARGE43.19}\\\LARGE0.994} & \makecell{\LARGE43.81\\\LARGE0.991} \\
\hline
\LARGE\textbf{Ours} & \LARGE18.91 & \makecell{\textbf{\LARGE43.35} \\ \textbf{\LARGE0.994}} & \makecell{\textbf{\LARGE47.16}\\\textbf{\LARGE0.997}} & \makecell{\textbf{\LARGE44.71}\\\textbf{\LARGE0.988}} & \makecell{\LARGE47.46\\\textbf{\LARGE0.996}} & \makecell{\LARGE41.75\\\textbf{\LARGE0.995}} & \makecell{\textbf{\LARGE45.35}\\\textbf{\LARGE0.997}} & \makecell{\textbf{\LARGE43.73}\\\textbf{\LARGE0.990}} & \makecell{\textbf{\LARGE43.98}\\\textbf{\LARGE0.996}} & \makecell{\textbf{\LARGE44.02}\\\textbf{\LARGE0.993}} & \makecell{\textbf{\LARGE43.19}\\\textbf{\LARGE0.997}} & \makecell{\textbf{\LARGE44.47}\\\textbf{\LARGE0.994}} \\
\hline
\end{tabular}}
\label{simu}
\end{table*}

\section{Experiments}
\label{sec:experiments}

\subsection{Experimental Settings}
\textbf{Dataset.} We used two simulated hyperspectral datasets: CAVE~\cite{yasuma2010generalized}, KAIST~\cite{choi2017high} and one real dataset~\cite{wang2016adaptive}. The CAVE dataset contains 32 hyperspectral images with a spatial resolution of 512$ \times $512 pixels. The KAIST dataset consists of 30 hyperspectral images with a higher spatial resolution of 2704$ \times $3306 pixels and a spectral dimension of 31. Following the protocol established in previous works~\cite{cai2022mask,huang2021deep,Wang_2024_CVPR}, we used the CAVE dataset for training set and select a subset of 10 scene crops from the KAIST dataset along with the real dataset for testing.

\textbf{Implementation Details.} We implemented our network on the PC with a single NVIDIA RTX 4090 GPU and built it in the PyTorch framework. In the training phase, the Adam optimizer~\cite{diederik2014adam} was used to optimize the model parameters. The initial learning rate was set to $4\times10^{-4}$ , and the learning rate was decayed using a cosine annealing schedule with a minimum value of $1\times 10^{-6}$. The batch size was set to 4. We cropped $256 \times 256$ patches from 3D cubes and input them into the network.

\subsection{Baseline Methods}
We compared our approach with three classic model-based spectral reconstruction methods (PFusion~\cite{he2021fast}, PIDS~\cite{chen2023prior} and TV~\cite{wang2015dual}), five end-to-end methods (BiSRNet~\cite{cai2024binarized}, CST~\cite{cai2022coarse}, HDNet~\cite{hu2022hdnet}, MST++~\cite{cai2022mst++} and SPECAT~\cite{yao2024specat}) and two depth unfolding methods (DAUHST~\cite{cai2022degradation}, In2SET~\cite{Wang_2024_CVPR}). 

\subsection{Metrics}
The reconstruction quality of hyperspectral images is evaluated using peak signal-to-noise ratio (PSNR) and structural similarity index (SSIM) metrics.

\subsection{Simulation and Real Data Results} 

\begin{wrapfigure}{r}{5.5cm}
  \vspace{-35pt} 
  \includegraphics[width=\linewidth]{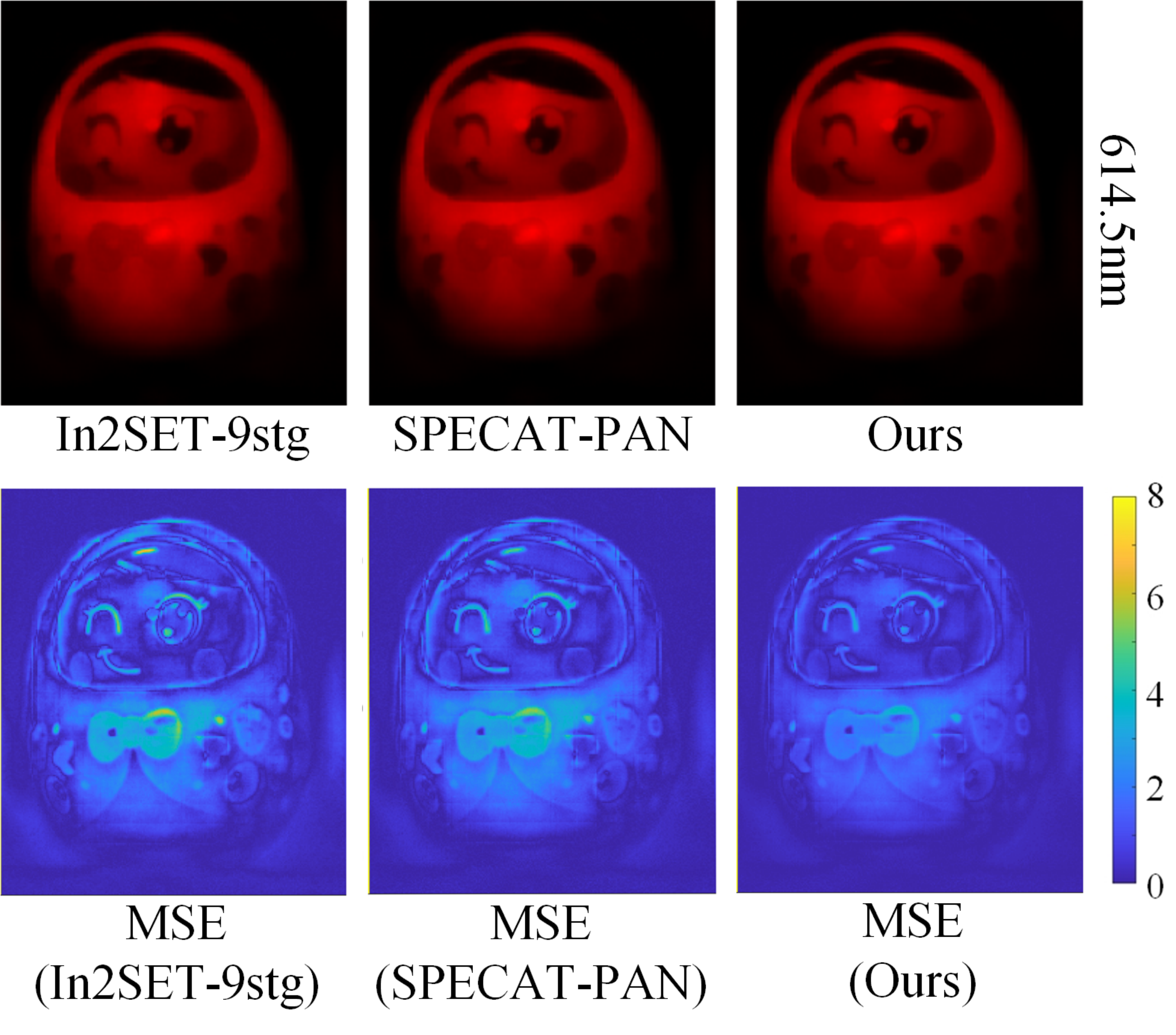}
  \caption{The visual comparisons between our method and SOTA methods on real dataset.}
  \vspace{-15pt} 
  \label{fig:real_result}
\end{wrapfigure}

\textbf{Numerical Results.} The metrics of different methods on ten simulated scenes are shown in Tab. \ref{simu}. Our method achieves superior performance in most scenes. The average PSNR and average SSIM of our method achieve 44.47 dB and 0.994, outperforming the second-best results by 0.66 dB and 0.003, respectively. Additionally, it can be observed that, compared to DAUHST-PAN-9stg and In2SET-9stg, PCMamba achieves higher reconstruction quality while requiring less than half of the computational cost.

\begin{figure*}[t]
  \centering
   \includegraphics[width=1\linewidth]{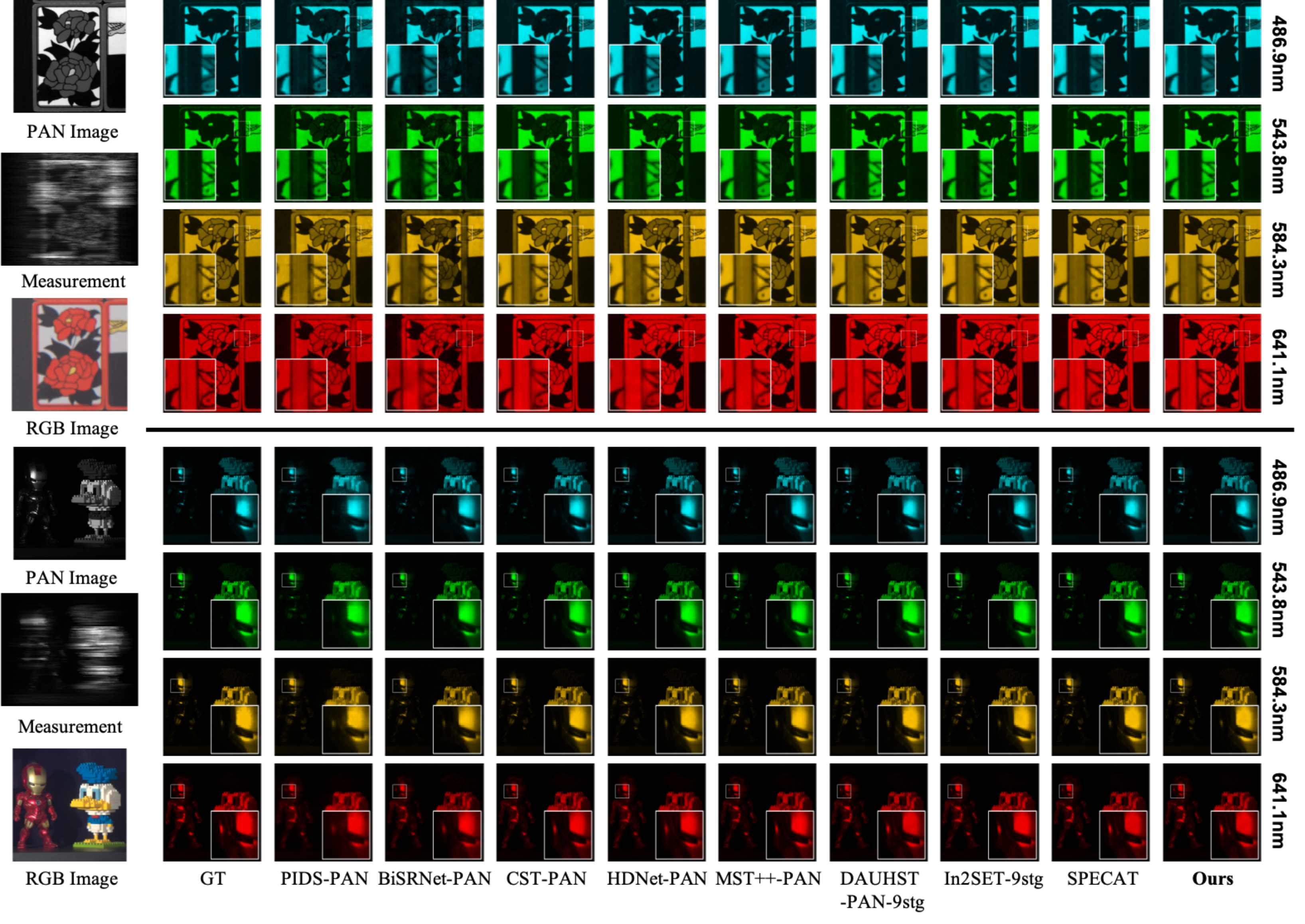}
   \caption{Comparison of the reconstruction results of different methods on two scenes from the KAIST dataset, including eight SOTA methods and our PCMamba. We select four bands (486.9 nm, 543.8 nm, 584.3 nm, and 641.1 nm) for visualization.}
   \label{fig:simu_result}
\end{figure*}

\textbf{Visual Results.} Fig. \ref{fig:real_result} shows the MSE residues between the reconstruction results and the ground truth on the real dataset, demonstrating that PCMamba achieves better spectral fidelity. To facilitate visual evaluation, we presented the reconstruction results of eight SOTA methods across four bands from the simulated dataset, alongside the ground truth. Fig. \ref{fig:simu_result} illustrates these reconstruction results. By zooming into local regions, it can be observed that our method reconstructs results that are closer to the ground truth. For example, the vertical line to the left of the bird in the first image is restored more sharply by our method compared to other SOTA methods. 

\subsection{Ablation Study}
TeX decomposition and the Cross-Modal Scanning Mamba Block (CSMB) are two key modules of PCMamba. We further conducted ablation experiments on simulated datasets to verify their effectiveness.

\begin{wraptable}[28]{r}{0.45\textwidth}  
\vspace{10pt}
\caption{Ablation studies on simulated dataset. `w/o' denotes without, and `w/' denotes with.}

\label{ablation_combined}
\centering

\begin{subtable}[t]{\linewidth}
    \caption{Ablation studies of the TeX decomposition.}
    \label{ablation_tex}
    \centering
    \begin{tabular}{lrr}
        \toprule
        TeX Process & PSNR$\uparrow$ & SSIM$\uparrow$ \\
        \midrule
        w/o TeX & 43.65 & 0.991 \\
        w/ TeX & \textbf{44.47} & \textbf{0.994} \\
        \bottomrule
    \end{tabular}
\end{subtable}

\vspace{15pt}

\begin{subtable}[t]{\linewidth}
    \caption{Ablation studies of the Cross-Scan.}
    \label{ablation_csr}
\centering
    \begin{tabular}{lrr}
        \toprule
        Cross-Scan & PSNR$\uparrow$ & SSIM$\uparrow$ \\
        \midrule
        w/o & 42.56 & 0.988 \\
        BFR = 0.3 & 44.03 & 0.993 \\
        BFR = 0.5 & 44.25 & 0.993 \\
        \textbf{BFR = 0.7} & \textbf{44.47} & \textbf{0.994} \\
        BFR = 0.8 & 44.36 & 0.993 \\
        BFR = 0.9 & 44.20 & 0.993 \\
        \bottomrule
    \end{tabular}
\end{subtable}

\vspace{15pt}

\begin{subtable}[t]{\linewidth}
    \caption{Ablation studies of the loss function terms.}
    \label{ablation_loss}
\centering
    \begin{tabular}{lrr}
        \toprule
        Config & PSNR$\uparrow$ & SSIM$\uparrow$ \\
        \midrule
        w/o $\mathcal{L}_{Rec}$ & 17.29 & 0.165 \\
        w/o $\mathcal{L}_{\mathcal{M}}$ & 44.08 & 0.993 \\
        Ours & \textbf{44.47} & \textbf{0.994} \\
        \bottomrule
    \end{tabular}
\end{subtable}

\end{wraptable}

\textbf{TeX Decomposition.}

As shown in Tab. \ref{ablation_combined} (a), we explored the impact of the introduced physical process on HSI reconstruction. Specifically, we removed the three decoders at the end of the U-net and directly generated the HSI. It is observed that the introduction of the TeX decomposition process improves PSNR by 0.82 dB, demonstrating that exploring the correlations between the latent physical properties within the input image is beneficial.

\textbf{Cross-Modal Scanning Mamba Block (CSMB).}

To validate the effectiveness of the cross-scanning scheme in CSMB, we replaced it with a Vanilla scan that performs pixel-wise scanning of both the backbone and PAN features. As shown in Tab. \ref{ablation_combined} (b), the cross-scanning scheme improved the PSNR by 1.91 dB, highlighting its effectiveness. Furthermore, we explored the impact of the backbone feature ratio (BFR) in the cross-scanning process. BFR is the proportion of backbone features in the encoder-decoder features. We found that maintaining an appropriate proportion between the backbone features and PAN features effectively enhanced the HSI reconstruction performance.

\textbf{Loss Function.}

We verified the effectiveness of each loss function by removing them individually, where the quantitative results are reported in Tab. \ref{ablation_combined} (c). It can be observed that the reconstruction loss $\mathcal{L}_{Rec}$ plays a major role, as it contains the primary supervisory information. On the other hand, the imaging process consistency loss $\mathcal{L}_{\mathcal{M}}$ effectively constrains the HSI generation process, resulting in an improvement in the quantitative metrics.

\section{Conclusion}
\label{sec:conclusion}
In this paper, we propose PCMamba, a physics-informed cross-modal SSM network for DCCHI. This is the first attempt to address the HSI reconstruction problem from the perspective of the physical process of spectral signal generation, aiming to provide theoretical guidance for future hyperspectral imaging tasks. PCMamba achieves the physical synthesis of HSI by exploiting three physical properties: temperature, emissivity, and texture. In addition, we design a Cross-Modal Scanning Mamba Block (CSMB), which learns more compact inter-modal inductive biases by performing cross-scanning without positional overlap across different modality features, thereby significantly reducing the computational cost of the SSM. Extensive experiments conducted on both real and simulated datasets demonstrate the effectiveness and efficiency of our method.


{\small
\bibliographystyle{plain}
\bibliography{main}
}

\end{document}